\documentstyle[12pt]{article}
\advance\textheight by \topskip
\begin{document}
\thispagestyle{empty}
\begin{flushright}
DF/IST- 1.96\\
November 1996
\end{flushright}
\vskip 8mm
\begin{center}
{\Large\bf Weak Lensing, Shear and the Cosmic Virial Theorem in a 
Model with a Scale-Dependent Gravitational Coupling}
\vskip 8mm
{\bf Orfeu Bertolami}\footnote{E-mail: orfeu@alfa.ist.utl.pt}
\vskip .1mm
{\it Instituto Superior T\'ecnico - Departamento de F\'\i sica}
\vskip 0.5mm
{Av. Rovisco Pais 1, 1096 Lisboa Codex, Portugal}
\end{center}
\vskip 10mm

\setlength{\baselineskip}{0.65cm}

{\centerline{\Large\bf Abstract} 

\begin{quotation}
\vskip -0.3cm
It is argued that, in models where the gravitational coupling is 
scale-dependent,
predictions concerning weak gravitational lensing and shear are essentially
similar to the ones derived from General Relativity. This is consistent 
with recent negative results of 
observations of the MS1224, CL2218 and A1689 systems aimimg to infer 
from those methods the 
presence of dark matter. It is shown, however, that 
the situation is quite different when  an analysis based on the 
Cosmic Virial Theorem is concerned. 
\end{quotation}

\newpage

It has been recently reported that studies of the distortion of the surface 
brightness of background faint galaxies due to the gravitational shear of
MS1224.7+2007 are consistent with an  excess of mass of about a factor 3 in
relation to the one inferred from the mass-to-light ratio analysis 
\cite{Fahlman}. This excess has been attributed to the presence of dark matter.
Distortion effects of this nature were also reported for the
CL 0024 system \cite{Bonnet} and similar studies were performed for the 
CL 1455 and CL 0016+16 systems \cite{Smail}. More recent studies indicate,
however, that discrepancies are much smaller and can be accounted for by 
internal dynamical effects other than the presence of dark matter. Furthermore,
it is found that for systems such as A 2218
the mass estimated by the weak gravitational lensing method is
consistent with the one inferred from the X-ray data \cite{Squires}, while in
different studies for the A2218, A1689 and A2163 systems, discrepancies
actually suggest that the inferred mass distribution 
is somewhat more compact than the one arising 
from studies of the X-ray profiles \cite{Miralda}.  
Thus, one is led to conclude that despite its advantages 
(independence from the virialization condition and from the hypothesis of
sphericity) over other dynamical methods, such as the study of 
rotation curves of 
galaxies and of the motion of galaxies in groups, the weak gravitational
lensing method (see eg. \cite{Waerbeke} for a critical assessment) has not 
been able, so far, to provide an unequivocal evidence of the presence of dark
matter.

In what follows we shall argue that these most recent results are consistent 
with what
one should expect from models where the presence of dark matter is replaced,
at least partially (see eg. \cite{BertBell}), by the dynamical effects 
associated with the scale-dependence of the gravitational coupling arising 
from various models of quantum gravity. 

In many approaches to quantum gravity the gravitational coupling  
turns out to have a momentum or scale dependence. In 1-loop  
higher-derivative quantum gravity this dependence arises as the  
theory has logarithmic divergences and the corresponding  
$\beta$-functions are negative meaning that 1-loop quantum gravity models  
are asymptotically free \cite{Julve}. Similar conclusions are drawn from
applying exact renormalization group techniques to gravity \cite{Eli, Reuter}.
This implies that the coupling constants of  
higher-derivative theories of gravity are actually 
momentum or scale-dependent.  
For an appropriate choice of the ``confining'' scale, the running of the  
coupling constants and in particular of the gravitational coupling  
can manifest itself macroscopically, as suggested in Ref.\cite{Goldman}. 
This behaviour is consistent with the absence of screening in gravity. 
Moreover, this implies that all classical equations depending on $G$ will 
also exhibit a dependence on scale. 

The cosmological implications of this feature of quantum gravity models
were discussed in Refs. \cite{Bertolami,Bertolami1} and  
compared with phenomenology in \cite{BertBell, BertBell1}. 
In  
particular, it was pointed out in the latter that the 
behaviour of the  
gravitational coupling with scale, as suggested in \cite{Goldman}, 
is consistent  
with known astrophysical and cosmological bounds
but still requires the presence of dark  
matter in the halo of spiral galaxies to explain the flatness 
of their rotation velocity curves, although that 
can be ensured with about 45\% less dark matter \cite{BertBell, Bottino}. 

Let us now turn to the discussion of the gravitational lensing phenomena. 
Gravitational lensing is based on the prediction of General Relativity that 
light passing at a distance $b$ from a spherical mass distribution is 
deflected by an angle that, in lowest order, is given by:
$$
\alpha =\frac{4GM(b)}{bc^2}\ ,\eqno(1)
$$
where $M(b)$ denotes the mass contained in $b$. The mass, $M(r)$, as 
a function of the distance from the centre can be inferred from 
Kepler's Third Law, $M(r) = r v^2/G$, where $v$ is the rotation velocity of a 
mass at an orbit of radius $r$ (see Ref. \cite{BertBell} for
a discussion in the context of scale-dependent gravitational coupling models).
Hence, assuming the lenses are spiral galaxies whose rotation velocity is 
constant for $r > 10-20~kpc$ and that $\rho(r) \propto r^{-2}$, then:
$$
\alpha = 2\pi \left(\frac{v}{c}\right)^2 \ ,\eqno(2)
$$
which is independent of the impact paramenter, $b$, and $G$ 
(see for instance \cite{Dar}). For typical values 
of the rotation velocity of spiral galaxies, $v \sim 250~km~s^{-1}$, 
one finds $\alpha \approx 1''$. Essentially similar results would follow
if elliptical galaxies are considered as the lenses instead, 
provided the rate of change of their total energy is unimportant, such that
one can still use the Virial Theorem to evaluate their
mean square velocity (see eq. (7) below).

Other lensing phenomena, such as the Einstein Rings, Crosses and Arcs are all 
given in terms of $\alpha$ \cite{Wu} and a ratio of geometrically 
relevant distances and
are therefore $G$-independent too. It then follows that one should 
not expect, at
least in lowest order, any scale dependence in the measured lensing 
parameters. The same can be said about the so-called shear, which for any of 
the models discussed in the literature (power law or de Vaucouleurs),
is $G$ independent \cite{Bonnet}. Hence, once again, one should not expect 
any scale dependence 
on this quantity as well. Since most recent observations
\cite{Squires} indicate that the distortion of the surface brightness
of background galaxies can be accounted for by internal dynamical effects
other than dark matter, and as the running of $G$ mimics the presence of 
dark matter, consistency with observations implies, as shown, 
that the scale dependence
of $G$ does not affect gravitational lensing phenomena.  
A scale dependence on the gravitational lensing phenomena can nevertheless 
be inferred,
if in order to estimate the relevant distances involved, the Hubble constant 
is used (see discussion in Ref. [7]) as this quantity is 
scale dependent (cf. eq. (9) below).

Of course, one should expect to draw conclusions about the presence 
of dark matter and/or about a possible scale dependence of $G$ from other 
dynamical studies, such
as the ones based on the Virial Theorem. Indeed, the
so-called Cosmic Virial Theorem is based on the assumption that particles 
with mass $m_i$ and position vectors $\vec x_i$ that are under the action of 
a gravitational potential between two particle densities $\rho (x_i)$  
given by
$$
W=-{1\over 2} {G\over M} a^5 (t) \int d^3 x_1 d^3 x_2 {[(\rho (x_1)  
-\rho_B) (\rho (x_2) -\rho_B)] \over x_{12}}\ , \eqno(3)
$$
where $M=\sum_i mi$, $a(t)$ is the scale factor of the Universe and  
$\rho_B$ an average background density, can be 
described by the Hamiltonian function
$$
{\cal H} = M (K+W)\eqno(4)
$$
with
$$
K\equiv {1\over 2} \frac{\sum m_i (a \dot{\vec x_i})^2}{M}\ .\eqno(5) 
$$
Furthermore, by conveniently writing $W$ in terms of the well known 
mass auto-correlation function $\xi (x)$,

$$
W = -{1\over 2} G \rho_B a^2 \int d^3 x {\xi (x)\over x}\ ,\eqno(6)
$$
one can obtain the Layzer-Irvine equation for the variation of the total 
energy \cite{Peebles}:
$$
{d\over dt} (K+W) + H (2K +W) = 0\ ,\eqno(7)
$$
where $H(t) \equiv \dot a(t)/a(t)$ is the Hubble parameter.

Equation (7) allows one to infer the rate of change of the total energy in  
terms of the Hubble parameter and the quantity 2K+W. The standard Virial 
Theorem readily follows from (7) if one neglects the rate of change of the 
total energy. Moreover, this equation also  
allows one to obtain an expression for the mass-weighted mean square  
velocity \cite{Peebles}
$$
\bar v^2 = 2\pi G \rho_B J_2 (x)\ ,\eqno(8)
$$
where $J_2 (x) = \int_0^x dx x \xi (x)$.

From the above discussion on the scale dependence of $G$, one has that the 
Friedmann equation, which describes the evolution of expansion rate of  
the Universe in terms of the matter energy density in an homogeneous   
and isotropic space-time should also, due to the presence of the gravitational
coupling, exhibit a scale dependence. Considering for simplicity a spatially 
flat Universe, one has \footnote{Notice that from this Friedmann
equation it implies that the Hubble constant has itself a scale dependence. 
As far as observations are concerned, this possibility cannot yet be 
ruled out before the discrepancy betweeen results obtained using 
as standard candles  
Type I supernovae (from which follows that 
$H_0 = (57\pm 4)\ {\rm km \ s}^{-1} {\rm Mpc}^{-1}$ \cite{Sandage}) 
and classical Cepheid variables (that yield 
$H_0 = (82 \pm 17)\ {\rm km \ s}^{-1} {\rm Mpc}^{-1}$ \cite{Freedman}) 
is fully understood. The recent discovery of a correlation between 
the peak luminosity and the luminosity decay of Type I supernovae 
does imply that values of the Hubble constant obtained 
using those stars as standard 
candles tend to become higher, 
$H_0 = (67 \pm 7)\ {\rm km \ s}^{-1} {\rm Mpc}^{-1}$ \cite{RPK}, opening up 
the possibility 
for a convergence between the two methods. This is the main thrust behind the 
Supernova Cosmology Project \cite{AGKim}. In case the Hubble 
constant is 
shown to have the same value at all scales, then one has either 
to give up the idea that $G$ has a scale dependence or instead 
the simple block renomalization procedure used here to generalize the 
classical equations
(or also to put the burden to explain the scale independence of $H$ on 
$\rho_B$).}:
$$
H^2{(l)} = {8 \pi G (l)\over 3} \rho_B \ .\eqno(9)
$$
It then follows that one can readily generalize  
eq. (7):
$$
{d\over dt} (K+W (l)) = -H (l) (2K+W(l))\ .\eqno(10)
$$
This generalization of the Layzer-Irvine equation indicates  
that the variation of the total energy, $K+W(l)$, is enhanced by the  
scale dependence of the rate of expansion $H(l)$ and by the $G(l)$ dependence 
in $W$. In order to estimate the effect of this dependence, we use the  
following fit for $G(l)$ in terms of the proper distance, $l$ 
\cite{Goldman} (see also  
Refs. \cite{BertBell,BertBell1,Kim}):
$$
G(l) = G\left[ 1+0.3 \left({l\over kpc}\right)^{0.15}\right] \ .\eqno(11)
$$
Thus, for a system like
MS 1224.7+2007 which stretches over $l= 2 Mpc$ [2] we get,  
$G (l=2 Mpc)= 1.94  G$, meaning that standard virial analysis of the 
system, based on eq. (8), underestimates the mass-weighted mean square velocity
by a factor 2.

In summary, we have seen that in models where the gravitational coupling is
scale dependent, predictions about the gravitational lensing and shear are,
at least to lowest order, essentially similar to those of General Relativity. 
This implies that the most recent negative results of the use of 
gravitational lensing methods 
to infer the presence of dark matter are consistent with 
our predictions.
We believe however, that the situation is, at cosmological scales, 
significantly different as the effective gravitational
coupling is in this case significantly greater than Newton's constant. 
That can be used to account
for features of the large scale structure of the Universe without
invoking large amounts of dark matter \cite{Bertolami}. 

\begin{flushleft}
{\bf Acknowledgements}

\noindent
It is pleasure to thank Juan Garc\'\i a-Bellido for the discussions 
which gave rise to this work and for the relevant subsequent
suggestions.
\end{flushleft}

\end{document}